\documentclass[aps]{revtex4}

\usepackage{amsmath,amssymb,amsfonts}
\usepackage{graphicx}
\usepackage{color}
\allowdisplaybreaks
\def\be{\begin{equation}}
\def\ee{\end{equation}}
\def\bea{\begin{eqnarray}}
\def\eea{\end{eqnarray}}
\def\ba{\bea}
\def\ea{\eea}
\def\bm{\bibitem}
\def\rw{\rightarrow}

\def\del{\partial}

\def\ra{\rangle}
\def\la{\langle}

\def\no{\nonumber}
\def\negs{\!\!\!\!\!\!\!\!}
 
\def\al{\alpha}
\def\bet{\beta}

\def\de{\delta}
\def\ep{\epsilon}

\def\sg{\sigma}
\def\th{\theta}
\def\om{\omega}
\def\Gm{\Gamma}
\def\De{\Delta}
\def\Lm{\Lambda}

\def\tht{\th}
     
\def\vk{\boldsymbol{k}}    
\def\vq{\boldsymbol{q}} 
\def\vl{\boldsymbol{l}} 
\def\vx{\boldsymbol{x}}     
    
\def\vph{\vec\phi}

\def\bld{\boldsymbol}
\def\bD{\bld{D}}

\def\bU{\boldsymbol{U}}

\def\bL{\boldsymbol{\Lambda}}

\def\ov{\overline} 

\def\oD{\overline{D}}

\begin{document}

\setcounter{page}{1}

\title{Evaluating transport coefficients in real time thermal field theory}

\author{S. \surname{Mallik}}
\email{mallik@theory.saha.ernet.in}
\affiliation{Theory Division, Saha Institute of Nuclear Physics, 1/AF
 Bidhannagar, Kolkata 700064, India}  
\author{Sourav \surname{Sarkar}}
\email{sourav@veccal.ernet.in}  
\affiliation{Theoretical Physics Division, Variable Energy Cyclotron
Centre, 1/AF, Bidhannagar, Kolkata, 700064, India}


\begin{abstract}
Transport coefficients in a hadronic gas have been calculated earlier in the 
{\em imaginary time} formulation of thermal field theory. The steps involved are 
to relate the defining {\em retarded} correlation function to the corresponding 
{\em time-ordered} one and to evaluate the latter in the conventional
perturbation expansion. Here we carry out both the steps in the {\em real time}
formulation.
\end{abstract}


\maketitle

\section{Introduction}
\setcounter{equation}{0}
\renewcommand{\theequation}{1.\arabic{equation}}

Thermal quantum field theory has been formulated in the imaginary as well
as real time \cite{Matsubara,Mills,Umezawa,Niemi,Kobes}. For time independent 
quantities such as the partition function, the imaginary time formulation is
well-suited and stands as the only simple method of calculation. However,
for time dependent quantities like two-point correlation functions, the use of 
this formulation requires a continuation to imaginary time and possibly back to
real time at the end. On the other hand, the real time formulation provides
a convenient framework to calculate such quantities, without requiring any
such continuation at all.

A difficulty with the real time formulation is, however, that all two-point 
functions take the form of $2\times 2$ matrices. But this difficulty is only 
apparent: Such matrices are always diagonalisable and it is the $11$- component of 
the diagonalised matrix that plays the role of the single function in the
imaginary time formulation. It is only in the calculation of this $11$-component 
to higher order in perturbation that the matrix structure appears in a non-trivial 
way.

In the literature transport coefficients are evaluated using the imaginary time 
formulation \cite{Hosoya,Lang,Jeon}. Such a coefficient is defined by the 
{\em retarded} correlation function of the components of the energy-momentum tensor. 
As the conventional perturbation theory applies only to {\em time-ordered}
correlation functions, it is first necessary to relate the two types of
correlation functions using the K\"{a}llen-Lehmann spectral representation
~\cite{Kallen,Lehmann,Fetter,MS}. We find this relation directly in real time 
formulation. The time-ordered correlation function is then calculated also in the 
covariant real time perturbative framework.

It suffices to illustrate the procedure with one transport coefficient, say
the shear viscosity. It is given by \cite{Zubarev,Hosoya,Lang},
\be
\eta=\frac{1}{10}\int_{-\infty}^0 dt_1 e^{\ep
t_1}\int_{-\infty}^{t_1}dt'\,i\int d^3x'\, \th (-t')
\la [\pi^{\alpha\beta}(0),\pi_{\alpha\beta}(\vx',t')]\ra\,,
\ee
where the space integral is over a retarded two-point function. Here $\la O\ra$ 
for any operator $O$ denotes {\em equilibrium} ensemble average,
\be
\la O\ra = {\rm Tr}(e^{-\bet H} O)/Z\,, ~~~~~~~~Z=Tre^{-\bet H}\,,
\ee
and $\pi^{\alpha\beta}(x)$ is the traceless part of the energy-momentum tensor. 
At temperatures sufficiently below phase transition, the pionic degrees of freedom 
dominate the hadron gas \cite{Gerber}; so one takes only their contribution to this 
tensor, getting
\be
\pi_{\alpha\beta}(x)=(\De_{\alpha}^{\rho}\De_{\beta}^{\sg}
-\frac{1}{3}\De_{\alpha\beta}\De^{\rho\sg})\del_\rho\vph (x)\cdot\del_\sg\vph (x)\,,
~~~~~~~~\De_{\al\bet}=g_{\al\bet}-u_\al u_\bet\,,
\ee
where $\vph (x)$ denotes the pion triplet and $u^\mu$ is the four-velocity
of the fluid, which is $(1,\vec{0})$ in the comoving frame. 

In Sec.~2 we derive the spectral representations for the retarded  and time-ordered 
correlation functions in the real time version of thermal field theory. The 
time-ordered function is then calculated to lowest order with complete propagators in 
Sec.~3. We conclude in Sec.~4. 
 
\section{Real-time formulation}
\setcounter{equation}{0}
\renewcommand{\theequation}{2.\arabic{equation}}

\begin{figure}
\centerline{\includegraphics[scale=0.5]{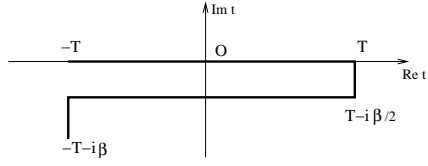}}
\caption{The contour $C$ in the complex time plane used here for the real
time formulation.}
\end{figure}

Here we review the real time formulation of thermal field theory leading to the
spectral representations of bosonic two-point functions \cite{MS}. This formulation 
begins with a comparison between the time evolution operator
$e^{-iH(t_2-t_1)}$ of quantum theory and the Boltzmann weight $e^{-\bet H}
=e^{-iH(\tau-i\bet-\tau)}$ of statistical physics, where we introduce $\tau$ as a 
complex variable. Thus while for the time evolution operator, the times $t_1$ and
$t_2$ $~(t_2 > t_1)$ are any two points on the real line, the Boltzmann
weight involves a path from $\tau$ to $\tau-i\bet$ in the complex time
plane. Setting this $\tau =-T$, where $T$ is real, positive and large, we
can get the contour $C$ shown in Fig.~1, lying within the region of analyticity
in this plane and accomodating real time correlation functions \cite{Mills,Niemi}.

Let a general bosonic interacting field in the Heisenberg representation be
denoted by $\Phi_l(x)$, whose subscript $l$ collects the index (or indices)
denoting the field component and derivatives acting on it. Although we shall
call its two-point function as propagator, $\Phi_l(x)$ can be an elementary 
field or a composite local operator. (If $\Phi_l(x)$ denotes the pion field, it 
will, of course, not have any index).

The thermal expectation value of the product $\Phi_l(x)\Phi^\dag_{l'}(x')$ may
be expressed as
\be
\la \Phi_l(x)\Phi^\dag_{l'}(x')\ra=\frac{1}{Z}\sum_{m,n} \la
m|\Phi_l(x)|n\ra\la n|\Phi^\dag_{l'}(x')|m\ra\,,
\ee
where we have two sums, one to evaluate the trace in eq.(1.2) and the other
to separate the field operators. They run over a complete set of states,
which we choose as eigenstates $|m\ra$ of four-momentum $P_\mu$. Using 
translational invariance of the field operator,
\be
\Phi_l(x)=e^{iP\cdot x}\Phi_l(0)e^{-iP\cdot x}\,,
\ee
we get
\be
\la\Phi_l(x)\Phi^\dag_{l'}(x')\ra=\frac{1}{Z}\sum_{m,n}e^{-\beta E_m}\,e^{i(k_m-k_n)\cdot (x-x')}
\la m|\Phi_l(0)|n\ra\la n|\Phi^\dag_{l'}(0)|m\ra\,.
\ee
Its spatial Fourier transform is
\bea
&&\negs \int d^3x\,e^{-i\vk\cdot(\vx-\vx')}\la\Phi_l(x)\Phi^\dag_{l'}(x')\ra\no\\
&&\negs =\frac{(2\pi)^3}{Z} \sum_{m,n}e^{-\beta E_m}\,e^{i(E_m-E_n)(\tau-\tau')}
\de^3(\vk_m-\vk_n+\vk)\la m|\Phi_l(0)|n\ra\la n|\Phi^\dag_{l'}(0)|m\ra\,,
\eea
where the times $\tau,\, \tau'$ are on the contour $C$. We now insert unity on the 
left of eq.~(2.4) in the form
\[1=\int_{-\infty}^\infty dk_0' \de(E_m-E_n+k_0')\,.\]
(We reserve $k_0$ for the variable conjugate to the real time.) Then it may be 
written as 
\be
\int d^3x \,e^{-i\vk\cdot(\vx-\vx')}\la\Phi_l(x)\Phi^\dag_{l'}(x')\ra
=\int\frac{dk_0'}{2\pi}e^{-ik_0'(\tau-\tau')}M^+_{ll'}(k_0',\vk)\,,
\ee
where the spectral function $M^+$ is given by $[k'_\mu=(k_0',\vk)]$
\be
M^+_{ll'}(k')=\frac{(2\pi)^4}{Z}\sum_{m,n}e^{-\beta E_m}\,\de^4(k_m-k_n+k')
\la m|\Phi_l(0)|n\ra\la n|\Phi^\dag_{l'}(0)|m\ra\,.
\ee

In just the same way, we can work out the Fourier transform of 
$\la\Phi^\dag_{l'}(x')\Phi_l(x)\ra$ 
\be
\int d^3x \,e^{-i\vk\cdot(\vx-\vx')}\la\Phi^\dag_{l'}(x')\Phi_l(x)\ra
=\int\frac{dk_0'}{2\pi}e^{-ik_0'(\tau-\tau')}M^-_{ll'}(k_0',\vk)\,,
\ee
with a second spectral function $M^-$ is given by
\be
M^-_{ll'}(k')=\frac{(2\pi)^4}{Z}\sum_{m,n}e^{-\beta E_m}\,\de^4(k_n-k_m+k')
\la m|\Phi^\dag_{l'}(0)|n\ra\la n|\Phi_l(0)|m\ra\,.
\ee
The two spectral functions are related by the KMS relation \cite{Kubo,Martin}
\be
M^+_{ll'}(k)=e^{\beta k_0}M_{ll'}^-(k)\,,
\ee
in momentum space, which may be obtained simply by interchanging the dummy indices 
$m,n$ in one of $M^\pm_{ll'}(k)$ and using the energy conserving $\de$-function.

We next introduce the {\it difference} of the two spectral functions,
\be
\rho_{ll'}(k) \equiv M_{ll'}^+(k)-M_{ll'}^-(k)\,,
\ee
and solve this identity and the KMS relation (2.9) for $M^\pm_{ll'}(k)$, 
\be
M^+_{ll'} (k)=\{1+f(k_0)\}\rho_{ll'} (k)\,, ~~~~ M^-_{ll'}(k)=f(k_0)\rho_{ll'}(k)\,,
\ee
where $f(k_0)$ is the distribution-like function
\be
f(k_0)=\frac{1}{e^{\beta k_0}-1}\,,~~~~~~~~-\infty <k_0 < \infty\,.
\ee
In terms of the true distribution function
\be 
n(|k_0|)=\frac{1}{e^{\beta |k_0|}-1}\,,
\ee
it may be expressed as
\bea
f(k_0)&=& f(k_0)\{\tht (k_0)+\tht(-k_0)\}\no\\
&=&  n\ep(k_0)-\tht (-k_0)\,.
\eea

With the above ingredients, we can build the spectral representations for the two 
types of thermal propagators. First consider the {\em time-ordered} one,
\bea
-iD_{ll'}(x,x')&=&\la T_c \Phi_l(x) \Phi^\dag_{l'}(x')\ra \no\\
&=&\tht_c(\tau-\tau')\la \Phi_l(x) \Phi^\dag_{l'}(x')\ra+\tht_c(\tau'-\tau)
\la \Phi^\dag_{l'}(x') \Phi_{l}(x)\ra\,.
\eea
Using eqs.~(2.5, 2.7, 2.11), we see that its spatial Fourier transform is given by
 \cite{Mills}
\be
D_{ll'}(\tau-\tau',\vk)=i\int_{-\infty}^\infty\frac{dk_0'}{2\pi}\rho_{ll'}(k_0',\vk)
e^{-ik_0'(\tau-\tau')}\{\tht_c(\tau-\tau')+f(k_0')\}\,,
\ee

As $T\rw \infty$, the contour of Fig.~1 simplifies, reducing essentially to
two parallel lines, one the real axis and the other shifted by $-i\bet/2$,
points on which will be denoted respectively by subscripts 1 and $2$, so
that $\tau_1=t,\, \tau_2=t-i\bet/2$ \cite{Niemi}. The propagator then consists of 
four pieces, which may be put in the form of a $2\times 2$ matrix.
The contour ordered $\th's$ may now be converted to the usual time ordered
ones. If $\tau,\tau'$ are both on line $1$ (the real axis), the $\tau$ and
$t$ orderings coincide, $\th_c(\tau_1-\tau'_1)=\th(t-t')$. If they are on
two different lines, the $\tau$ ordering is definite,
$\th_c(\tau_1-\tau'_2)=0,\, \th_c(\tau_2-\tau'_1)=1$. Finally if they are
both on line $2$, the two orderings are opposite,
$\th_c(\tau_2-\tau'_2)=\th(t'-t)$.

Back to real time, we can work out the usual temporal Fourier transform of
the components of the matrix to get
\be   
\bD_{ll'} (k_0,\vk)=\int_{-\infty}^{\infty}\frac{dk_0'}{2\pi}\rho_{ll'}(k_0',\vk)
\bL (k_0',k_0)\,,  
\ee
where the elements of the matrix $\bL$ are given by \cite{MS}
\bea
&& \Lm_{11}=-\Lm_{22}^* =\frac{1}{k_0'-k_0-i\eta}+2\pi if(k_0')\de(k_0'-k_0)\,,
\no\\
&& \Lm_{12}=\Lm_{21}=2\pi ie^{\beta k_0'/2}f(k_0')\de(k_0'-k_0)\,.
\eea
Using relation (2.14), we may rewrite (2.18) in terms of $n$, 
\bea
&& \Lm_{11}=-\Lm_{22}^* =\frac{1}{k_0'-k_0-i\eta\ep(k_0)}+2\pi in\ep(k_0)
\de(k_0'-k_0)\,,\no\\
&& \Lm_{12}=\Lm_{21}=2\pi i\sqrt{n(1+n)}\ep(k_0)\de(k_0'-k_0)\,.
\eea

The matrix $\bL$ and hence the propagator $\bD_{ll'}$ can be diagonalised
to give 
\be
\bD_{ll'}(k_0,\vk)=\bU 
\left(\begin{array}{cc}\oD_{ll'} & 0\\0 &
-\oD_{ll'}^{\,*}\end{array}\right)\bU\,,
\ee
where $\oD_{ll'}$ and $\bU$ are given by
\be
\oD_{ll'} (k_0,\vk)=\int_{-\infty}^{\infty}
\frac{dk_0'}{2\pi}\frac{\rho_{ll'}(k_0',\vk)}{k_0'-k_0-i\eta\ep(k_0)}\,,~~~~~~
\bU = \left( \begin{array}{cc} \sqrt{1+n} & \sqrt{n}\\\sqrt{n} &
\sqrt{1+n}\end{array}\right)\,.
\ee 
Eq.~(2.20) shows that $\oD$ can be obtained from any of the elements of the
matrix $\bD$, say $D_{11}$. Omitting the indices $ll'$, we get
\be
{\rm Re}\oD={\rm Re D}_{11}\,,~~~~{\rm Im}\oD=\tanh(\bet|k_0|/2){\rm Im D}_{11}\,.
\ee

Looking back at the spectral functions $M^{\pm}_{ll'}$  defined by (2.6, 2,8), we can 
express them as usual four-dimensional Fourier transforms of ensemble average of 
the operator products, so that $\rho_{ll'}$ is the Fourier transform of that of the 
commutator,
\be
\rho_{ll'}(k_0,\vk)=\int d^4y e^{ik\cdot (y-y')}\la
[\Phi_l(y),\Phi_{l'}(y')]\ra\,,
\ee
where the time components of $y$ and $y'$ are on the real axis in the
$\tau$-plane. Taking the spectral function for the free scalar field,
\be
\rho_0=2\pi\ep(k_0)\de (k^2-m^2)\,,
\ee
we see that $\oD$ becomes the free propagator, $\oD (k_0,\vk)=-1/(k^2-m^2)$.

We next consider the {\em retarded} thermal propagator
\be
D^R_{ll'}(x.x')=i\th_c(\tau-\tau')\la[\Phi_l(\vx,\tau),\Phi_{l'}(\vx',\tau')]\ra\,,
\ee
where again $\tau\,,\tau'$ are on the contour $C$ (Fig.~1). Noting
eqs.~(2.5, 2.7, 2.10) the three dimensional Fourier transform may immediately
be written as 
\be
D^R_{ll'}(\tau-\tau', \vk)=i\th_c(\tau-\tau')\int_{-\infty}^\infty
\frac{dk_0'}{2\pi}e^{-ik_0'(\tau-\tau')}\rho_{ll'} (k_0',\vk)\,.
\ee
As before we isolate the different components with real times and take 
the Fourier transform with respect to real time. Thus for the $11$-component we 
simply have
\be
D^R_{ll'}(t-t',\vk)_{11}=i\th (t-t')\int_{-\infty}^\infty
\frac{dk_0'}{2\pi}e^{-ik_0'(t-t')}\rho_{ll'} (k_0',\vk)\,,
\ee
whose temporal Fourier transform gives 
\be
D^R_{ll'}(k_0,\vk)_{11}=\int_{-\infty}^\infty             
\frac{dk_0'}{2\pi}\frac{\rho_{ll'} (k_0',\vk)}{k_0'-k_0-i\eta}\,.      
\ee
This $11$-component suffices for us, but we also display the complete matrix,
\be
\bD^R_{ll'}(k_0,\vk)= \left(\begin{array}{cc} D^R_{ll'}(k)_{11} & 0\\
\rho_{ll'}(k) \{\sqrt{\frac{n}{n+1}}\th(k_0)+\sqrt{\frac{n+1}{n}}\th(-k_0)\} &
-D^{R*}_{ll'}(k)_{11}\end{array}\right)\,.
\ee
Though we deal with matrices in real time formulation, it is the $11$-component
that is physical. Eqs.~(2.21) and (2.28) then show that we can continue the
{\em time-ordered} two-point function into the {\em retarded} one by simply 
changing the $i\ep$ prescription,
\be
D^R_{ll'}(k_0+i\eta,\vk)_{11}=\ov D_{ll'}(k_0+i\eta\ep(q_0)\to k_0+i\eta,\vk)\,.
\ee 
The point to note here is that for the time-orderd propagator, it is the
{\em diagonalised} matrix and not the matrix itself, whose $11$-component can be
continued in a simple way.

\section{Perturbative evaluation}
\setcounter{equation}{0}
\renewcommand{\theequation}{3.\arabic{equation}}

\begin{figure}
\centerline{\includegraphics[scale=0.5]{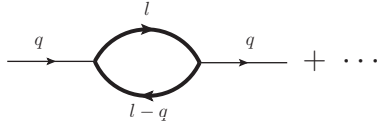}}
\caption{The first term in the so-called skeleton expansion of the two-point function.
 Heavy lines denote full propagators.}
\end{figure}

Clearly the spectral forms and their inter-relations derived above hold also
for the two-point function appearing in eq.~(1.1) for the shear viscosity. We
begin with four-dimensional Fourier transforms. To calculate the $11$-element of 
the the retarded two-point function
\be
\Pi_{11}^R(q)=i\int d^4x e^{iq(x-x')}\th(t-t')\la 
[\pi_{\alpha\beta}(\vx,t),\pi^{\alpha\beta}(\vx',t')]\ra\,,
\ee
we consider the corresponding time-ordered one,
\be
\Pi_{11}(q)=i\int d^4x e^{iq(x-x')}\la T
\pi_{\alpha\beta}(\vx,t),\pi^{\alpha\beta}(\vx',t')\ra\,,
\ee
which can be calculated perturbatively. To leading order, it is given by Wick 
contractions of pion fields in $\pi_{\alpha\beta}$ given by eq.~(1.3). 
In the so-called skeleton expansion, these contractions are expressed in terms of 
complete propagators (see Fig.~2) to get,
\be
\Pi_{11}(q)=i\int \frac{d^4l}{(2\pi)^4}N(l,q)D_{11}(l)D_{11}(l-q)\,,
\ee
where $N(l,q)$ is determined by the derivatives acting on the pion fields,
\be
N(l,q)=-6\,[\vl^2(\vl-\vq)^2+\frac{1}{3}\{\vl\cdot(\vl-\vq)\}^2].
\ee
To work out the $l_0$ integral in eq.~(3.3), it is more convenient to use $\Lm_{11}$
given by eq.~(2.18) than by eq.~(2.19). Closing the contour in the upper or lower 
half $l_0$-plane we get
\be
\Pi_{11}(q)=\int \frac{d^3l}{(2\pi)^3}N(\vl,\vq)\int\frac{dk_0'}{2\pi}
\rho(k_0',\vl)\frac{dk_0''}{2\pi}\rho(k_0'',\vl-\vq) K(q_0,k_0',k_0'')\,,
\ee
where 
\be
K=\frac{\{1+f(k_0')\}f(k_0'')}{q_0-(k_0'-k_0'')+i\eta}-
\frac{f(k_0')\{1+f(k_0'')\}}{q_0-(k_0'-k_0'')-i\eta}\,.
\ee
The imaginary part of $\Pi_{11}$ arises from the factor $K$,
\ba
{\rm Im}K&=&-\pi\left[\{1+f(k_0')\}f(k_0'')+f(k_0')\{1+f(k_0'')\}\right]
\de(q_0-(k_0'-k_0''))\nonumber\\
&=&-\pi\coth(\beta q_0/2)\{f(k_0'')-f(k_0')\}\de(q_0-(k_0'-k_0''))\,,
\ea
while its real part is given by the principal value integrals.

Having obtained the real and imaginary parts of $\Pi_{11}(q)$, we use
relations similar to eq.~(2.22) to build the $11$-element of the
diagonalised $\Pi$ matrix,
\be
\ov\Pi=\int \frac{d^3l}{(2\pi)^3}N(\vl,\vq)\int \frac{dk_0'}{2\pi}\rho(k_0',\vl)
\int \frac{dk_0''}{2\pi}\rho(k_0'',\vl-\vq)
\frac{\{1+f(k_0')\}f(k_0'')-f(k_0')\{1+f(k_0'')\}}
{q_0-(k_0'-k_0'')+i\eta\ep(q_0)}\,.
\ee
Finally $\ov\Pi$ can be continued to $\Pi_{11}^R$ by a relation similar to
eq.~(2.30), 
\be
\Pi^R_{11}=\int\frac{d^3l}{(2\pi)^3}N(\vl,\vq)\int \frac{dk_0'}{2\pi}\rho(k_0',\vl)
\frac{dk_0''}{2\pi}\rho(k_0'',\vl-\vq)
\frac{\{1+f(k_0')\}f(k_0'')-f(k_0')\{1+f(k_0'')\}}
{q_0-(k_0'-k_0'')+i\eta}\,.
\ee
Note that in eqs.~(3.8,3.9) we retain the $f(k_0')f(k_0'')$ terms in the
numerator to put it in a more convenient form. Change the signs of $k_0'$ and $k_0''$ 
in the first and second term respectively. Noting relations like $1+f(-k_0)=-f(k_0)$ 
and $\rho(-k_0)=-\rho(k_0)$ we get
\be
\Pi^R_{11}(q)=\int\frac{d^3l}{(2\pi)^3}N(\vl,\vq)\int\frac{dk_0'}{2\pi}\frac{dk_0''}{2\pi}
\rho(k_0',\vl)\rho(k_0'',\vl-\vq)f(k_0')f(k_0'')W(q_0,k_0'+k_0'')\,,
\ee
where
\be
W=\frac{1}{q_0+k_0'+k_0''+i\eta}-\frac{1}{q_0-(k_0'+k_0'')+i\eta}\,.
\ee

Returning to the expression (1.1) for $\eta$, we now get the three-dimensional
spatial integral of the retarded correlation function by setting $\vq =0$ in
eq.~(3.1) and Fourier inverting with respect to $q_0$,
\be
i\int d^3x'\,\th(-t')\la [\pi^{\alpha\beta}(\vec 0,0),\pi_{\alpha\beta}(\vx',t')]\ra
=-\int dq_0\,e^{iq_0t'}\Pi^R_{11}(q_0,\vq=0)\,.
\ee
This completes our use of the real time formulation to get the required result.
The integrals appearing in the expression for $\eta$ have been evaluated in Refs. 
\cite{Hosoya,Lang}, which we describe below for completeness.

As shown in Ref. \cite{Hosoya}, the integral over $t_1,\,t'$ and $q_0$ in
eqs.~(1.1) and (3.12) may be carried out trivially to give
\be
\eta=\left.\frac{i}{10}\frac{d}{dq_0}\Pi^R_{11}(q_0)\right|_{q_0=0}\,.
\ee
The $q_0$ dependence of $\Pi^R_{11}$ is contained entirely in $W$,
\be
\left.\frac{d}{dq_0}W(q_0)\right|_{q_0=0}=-\frac{1}{(k_0'+k_0''-i\eta)^2}+
\frac{1}{(k_0'+k_0''+i\eta)^2}=2\pi i\de'(k_0'+k_0'')\,.
\ee
Changing the integration variables in eq.~(3.10) from $k_0'$, $k_0''$ to 
$\ov k_0=k_0'+k_0''$ and $k_0=\frac{1}{2}(k_0'-k_0'')$ we get
\be
\eta=\int \frac{d^3l}{(2\pi)^3}N(\vl)\int \frac{dk_0}{(2\pi)^2}F(k_0,\vl)\,,
\ee
where
\be
F(k_0,\vl)=\left.\frac{d}{d\ov k_0}\left\{\rho\!
\left(\frac{\ov k_0}{2}+k_0,\vl\right)\rho\!\left(\frac{\ov k_0}{2}-k_0,\vl\right)
f\!\left(\frac{\ov k_0}{2}+k_0\right)f\!\left(\frac{\ov
k_0}{2}-k_0\right)\right\}\right|_{\ov k_0=0}\,.
\ee

It turns out that the integral over $k_0$ becomes undefined, if we try to evaluate
$F(k_0)$ with the free spectral function $\rho_0(k)$ given by eq.~(2.24).
As pointed out in Ref.~\cite{Hosoya}, we have to take the spectral function
for the complete propagator that includes the self-energy of the pion,
leading to its finite width $\Gm$ in the medium,
\be
\rho(k_0,\vl)=\frac{1}{i}\left[\frac{1}{(k_0-i\Gm)^2-\om^2}
-\frac{1}{(k_0+i\Gm)^2-\om^2}\right]\,,~~~~~~~~~~\om=\sqrt{\vl^2+m^2}\,.
\ee
Then $F(k_0,\vl)$ becomes
\be
F=-8\frac{k_0^2e^{\bet k_0}}{(e^{\bet k_0}-1)^2}
\frac{\bet\Gm^2}{\{(k_0-i\Gm)^2-\om^2\}^2\{(k_0+i\Gm)^2-\om^2\}^2}\,,
\ee
having double poles at $k_0=2\pi in/\bet$ for $n=\pm 1, \pm 2, \cdots$ and
also at $k_0=\pm\om\pm i\Gm$. The integral over $k_0$ may now be evaluated by
closing the contour in the upper/lower half-plane to get
\be
\int^{+\infty}_{-\infty}\frac{dk_0}{(2\pi)^2}
F(k_0,\vl)=-\frac{1}{8\pi}\frac{\bet}{\om^2\Gm}n(\om)\{1+n(\om)\}\,,
\ee
where we retain only the leading (singular) term for small $\Gm$. In this
approximation eq.~(3.15) gives~\cite{Lang}
\be
\eta=\frac{\beta}{10\pi^2}\int_0^\infty dl\,
l^6\frac{n(\om)\{1+n(\om)\}}{\om^2\Gm}\,.
\ee
The width $\Gm (l)$ at different temperatures is known \cite{Goity} from chiral 
perturbation theory \cite{Gasser}, using which $\eta$ has been evaluated numerically 
~\cite{Lang}.

\section{Conclusion}

Here we calculate a transport coefficient in the real time version of thermal field 
theory. It is simpler to the imaginary version in that we do not have to continue to 
imaginary time at any stage of the calculation. As an element in the theory of linear 
response, a transport coefficient is defined in terms of a retarded thermal two-point 
function of the components of the energy-momentum tensor. We derive K\"{a}llen-Lehmann
representation for any (bosonic) two-point function of both time-ordered and
retarded types to get the relation between them. Once this relation is
obtained, we can calculate the retarded function in the Feynman-Dyson
framework of the perturbation theory. 

Clearly the method is not restricted to transport coefficients. Any linear
response leads to a retarded two-point function, which can be calculated in
this way. Also quadratic response formulae have been derived in the real
time formulation \cite{Carrington}.

\section*{Acknowledgement}

One of us (S.M.) acknowledges support from Department of Science and
Technology, Government of India.

\end{document}